\documentstyle[preprint,aps]{revtex}

\begin{document}
\draft

 \input epsf

\author{B. Zheng, M. Schulz, and S. Trimper}

\address{Fachbereich Physik, Martin--Luther--Universit\"at
D -- 06099 Halle Germany}

\title{Monte Carlo Simulations of a Generalized n--spin facilitated kinetic
Ising Model}

\date{\today}

\maketitle

\begin{abstract}

A kinetic Ising model is analyzed where spin variables correspond to 
lattice cells with mobile or immobile particles. Introducing 
additional restrictions for the flip processes according to 
the n-spin facilitated kinetic Ising model and using Monte Carlo methods 
we study the freezing 
process under the influence of an additional nearest-neighbor interaction. 
The stretched exponential decay of the auto-correlation function is observed 
and the exponent $\gamma$ as well as the relaxation time are determined 
depending on the activation energy $h$ and the short range coupling $J$. The 
magnetization corresponding to the density of immobile particles is 
found to be the controlling parameter for the dynamic evolution.
\\*[3cm]
 
\end{abstract}

\section{introduction}
There is a continuous effort in describing of supercooled liquids 
using different approaches \cite{Goetze1,Goetze2,Jaeckle,Leutheusser}. 
However the phenomenon is generally not complete understood. 
Supercooled fluids reveal normally a stretched 
exponential decay of typical (e.g. density--density) correlation functions
and a non-Arrhenius behavior of the associated relaxation times. 
This slowing down in the dynamical behavior can be illustrated by a strongly
curved trajectory in the Arrhenius plot (relaxation time $\tau $ versus the
inverse temperature $T^{-1}$), empirically described by the well known
Williams--Landel--Ferry (WLF) relation \cite{WLF}. But in contrast to conventional 
phase transitions a long range order is not developed.\\ 
The characteristic slowing down of the dynamics is usually explained 
by an increasing cooperativity of local processes with decreasing temperature 
\cite{Adams}. This behavior is an universal phenomena of the glass 
transition.\\ 
Mode coupling theories \cite{Goetze1,Goetze3,Goetze4} (MCT) predict the 
existence of an ergodic behavior above a critical temperature $T_c$ and a 
nonergodic behavior below $T_c$. Note that $T_c$ is in the range between the 
melting temperature $T_m$ and the glass temperature $T_g$, i.e. $T_m>T_c>T_g$. 
At $T_c$ the system undergoes a sharp transition from an ergodic state to a
state with partially frozen (density) fluctuations. The slow 
$\alpha$--process within the MCT is thought to correspond to the actual 
dynamic glass transition whereas the fast $\beta $--process is often 
identified with a cage rattling or the boson peak.\\ 
Actually, the nonergodic state 
obtained from the original MCT below $T_c$ are approximately stable only for
a finite time interval. Strongly cooperative processes lead to a slow decay
of apparently frozen structures. This slow decay shows the typical above
mentioned properties corresponding to the dynamics of the main glass
transition (WLF like behavior of the relaxation time, stretched exponential
decay of the correlation function). This effect can be partially described
in terms of an extended mode coupling theory \cite{Goetze2,Goetze3}
introducing additional hopping processes.\\ 
There exist also various alternative descriptions \cite{Jaeckle,Fredrickson1}
which explain the cooperative motion of the particles inside a supercooled
liquid below $T_c$. One of these possibilities is the spin facilitated Ising
model \cite{Fredrickson1,Fredrickson2,Fredrickson3,Fredrickson4}, originally
introduced by Fredrickson and Andersen. The basic idea of these models
consists of a coarse graining of space and time scales and simultaneously a
reduction of the degrees of freedom. In details that means:
\begin{enumerate}
\item  {\it Coarse graining of spatial scales}: The supercooled liquid is
divided into cells in such a way that each cell contains a sufficiently
large number of particles which realize a representative number of molecular
motions. 
\item  {\it Reduction of the degrees of freedom}: Each cell will be
characterized by only one degree of freedom, i.e. the cell structure
enables us to attach to each cell an observable $s_j$ (usually denoted
as spin) which characterizes the actual dynamic state of particles inside
the cell $j$. The usual realization is given by the local density $\rho _j$
(particles per cell) with $s_j = 1$ if $\rho _j>\bar \rho $ and 
$s_j=-1$ if $\rho _j<\bar \rho $ where $\bar \rho $ is the averaged 
density of the system. 
This mapping implies consequently different mobilities of the
particles inside such a cell, i.e. $s_j=1$ corresponds to the immobile
solid like state and $s _j=-1$ to the mobile state of cell $j$. The set
of all spin observables forms a configuration, the time expansion of the 
corresponding probability distribution obeys a master equation.
\item  {\it Coarse graining of the time scale}: This step bases on the
assumption that fast processes (e.g. the $\beta$--process) are well 
separated from the slow $\alpha$--process. Hence, the original Liouville
equation of the supercooled liquid can be projected onto a simple
master equation without any memory terms. Therefore, the spin facilitated
kinetic Ising model is suitable for a description of a supercooled
liquid well below $T_c$ within the MCT and for sufficiently large time
scales.
\end{enumerate}
To make the time evolution of the glass configurations more transparent we
use the argumentation following the idea of Fredrickson and Andersen 
\cite{Fredrickson1,Fredrickson2,Fredrickson3,Fredrickson4}, 
i.e. we suppose that  
the basic dynamics is a simple process $s_j=+1\leftrightarrow s_j=-1$ 
controlled by the thermodynamical Gibb's measure and by self-induced
topological restrictions. In particular, an elementary flip at a given cell
is allowed only if the number of the nearest neighbored mobile cells 
($s_j=-1$) is equal or larger than a restriction number $f$ with $0<f<z$
($z$: coordination number). So, elementary flip processes and geometrical
restrictions lead to the cooperative rearrangement of the underlying system
and therefore to a mesoscopic model describing a supercooled liquid
below $T_c$. Such models 
\cite{Fredrickson1,Fredrickson2,Fredrickson3,Fredrickson4} are denoted as 
$f$--spin facilitated Ising model on a $d$--dimensional lattice, 
SFM$\left[ f,d\right]$. 
The SFM$\left[ f,d\right] $ can be classified as an Ising-like
model the kinetics of which is confined by restrictions of the ordering of
nearest neighbors to a given lattice cell. This self--adapting environments
influence in particular the long time behavior of the spin-spin and
therefore of the corresponding density-density correlation functions. 
These
models were studied numerically \cite{Schulz1,Schulz2,Schulz3,Harrowell} 
(SFM$\left[ 2,2\right] $) and recently also analytically 
\cite{Schulz4} (SFM$\left[ f,d\right] $).\\ 
For the present investigations, we generalize the usual 
SFM$\left[ f,d\right] $ by introduction of 
additional short range interactions which
favor (antiferromagnetic case) or prevent (ferromagnetic case) the formation 
of liquid--solid (mobile--immobile) interfaces. 
It is the aim to study a two--dimensional generalized 
SFM$\left[ 2,2\right]$ using Monte--Carlo simulations.

\section{model}

We consider a generalized spin facilitated Ising model with nearest-neighbor
interactions in two dimensions. The Hamiltonian of the model is the same as 
that of the standard two-dimensional Ising model with an external field
\begin{equation}
H= - h (J\  \sum_{<ij>}  s_i\ s_j + \sum_{i}  s_i)\;,\qquad s_i=\pm 1.
\label{esho10}
\end{equation}
In our notation, the inverse temperature $T$ and the Boltzmann
constant $k$ has been absorbed into the field $h$. Physically, the field 
corresponds to the difference of the energy per cell in the liquid and 
the solid state.   
In our later discussions, for convenience, we simply denote
$h=1/T$. Here
the coupling constant $J$ describes the nearest-neighbor 
interactions. In case of $J=0$, the original Fredrickson model
is recovered. As above mentioned, the dynamic evolution of 
the generalized SFM$\left[ 2,2\right] $
is subjected to a topological constraint that 
a spin flip is only possible if
\begin{equation}
 \frac {1}{2} \sum_{i} (1+ s_i) \leq f.
\label{esho20}
\end{equation}
In our simulations, the Metropolis algorithm is used
and $f$ takes its typical value $f=d$ with $d$ being the space dimension, here 
we chose $d=2$. To assure that the system evolves into the physical section
of the phase space, the initial configuration
is always taken to be $s_i \equiv -1$ which means that we start from the 
complete liquid--like state. 
After the system has reached its equilibrium,
we measure the auto-correlation function
\begin{equation}
 A(t) = \frac{1}{L^d}\, <\sum_i s_i(t^\prime) s_i(t+t^\prime)>
\label{esho30}
\end{equation}
with $L$ being the lattice size.
Practically an average over $t^\prime$ is made
in the numerical measurements. The lattice sizes are taken to be
$L=50$ or $L=100$ depending on $h$ and $J$.
Up to the time regime of our simulations, no visible  
finite size effect has been observed.
To achieve reliable results and estimated statistical errors, 
we have performed five runs of simulations. 
Total samples for average range from $50\ 000$ to
$500\ 000$. More samples are for larger values of $h$ and/or $J$.

\section{results and discussion}
\begin{figure}[h]
\epsfysize=6.5cm
\epsfclipoff
\fboxsep=0pt
\setlength{\unitlength}{0.6cm}
\begin{picture}(9,9)(0,0)
\put(-1.,-0.5){{\epsffile{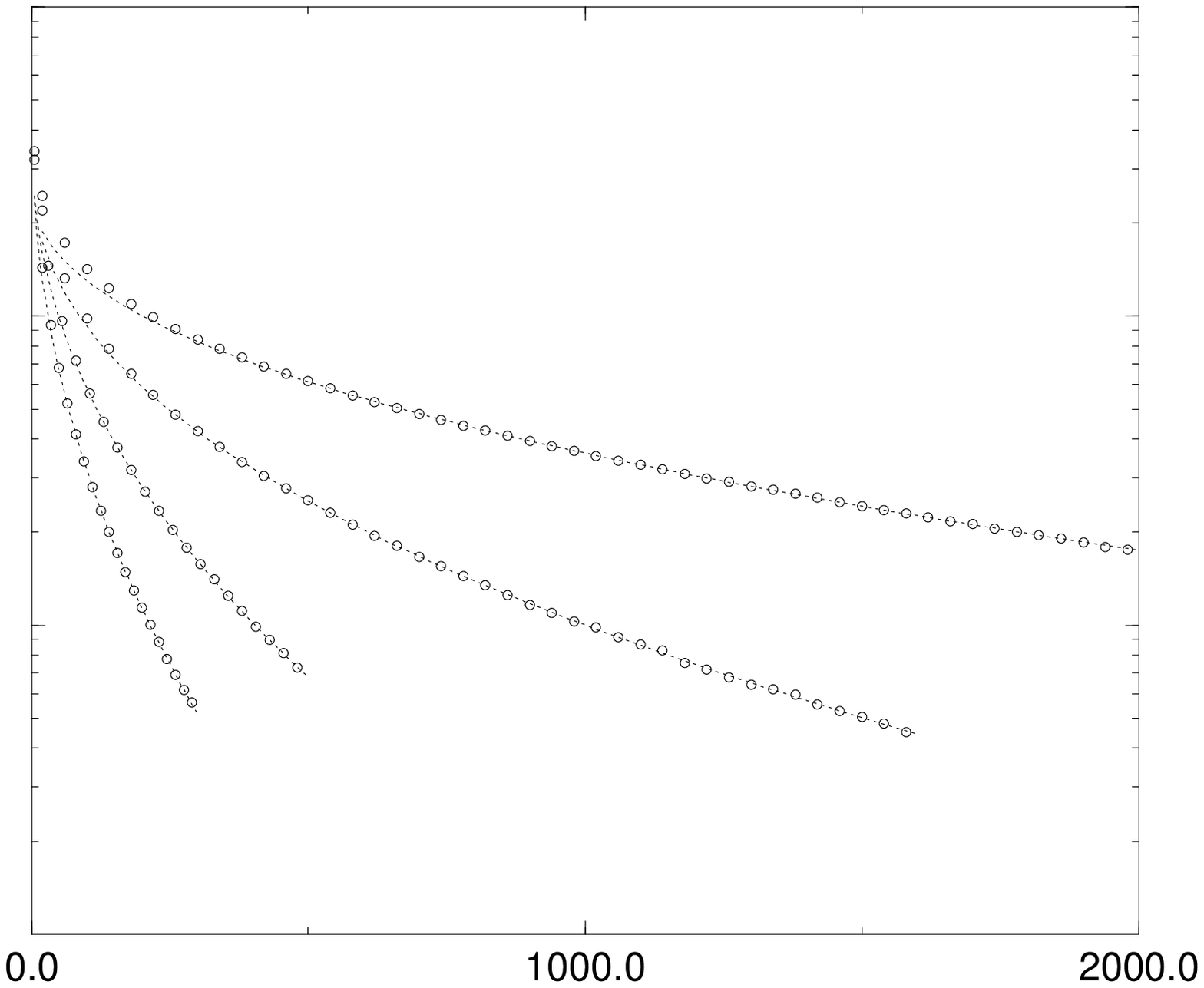}}}
\put(0.0,8){\makebox(0,0){\footnotesize $A(t)$}}
\put(8.0,0.5){\makebox(0,0){\footnotesize $t$}}
\epsfysize=6.5cm
\put(12.,-0.5){{\epsffile{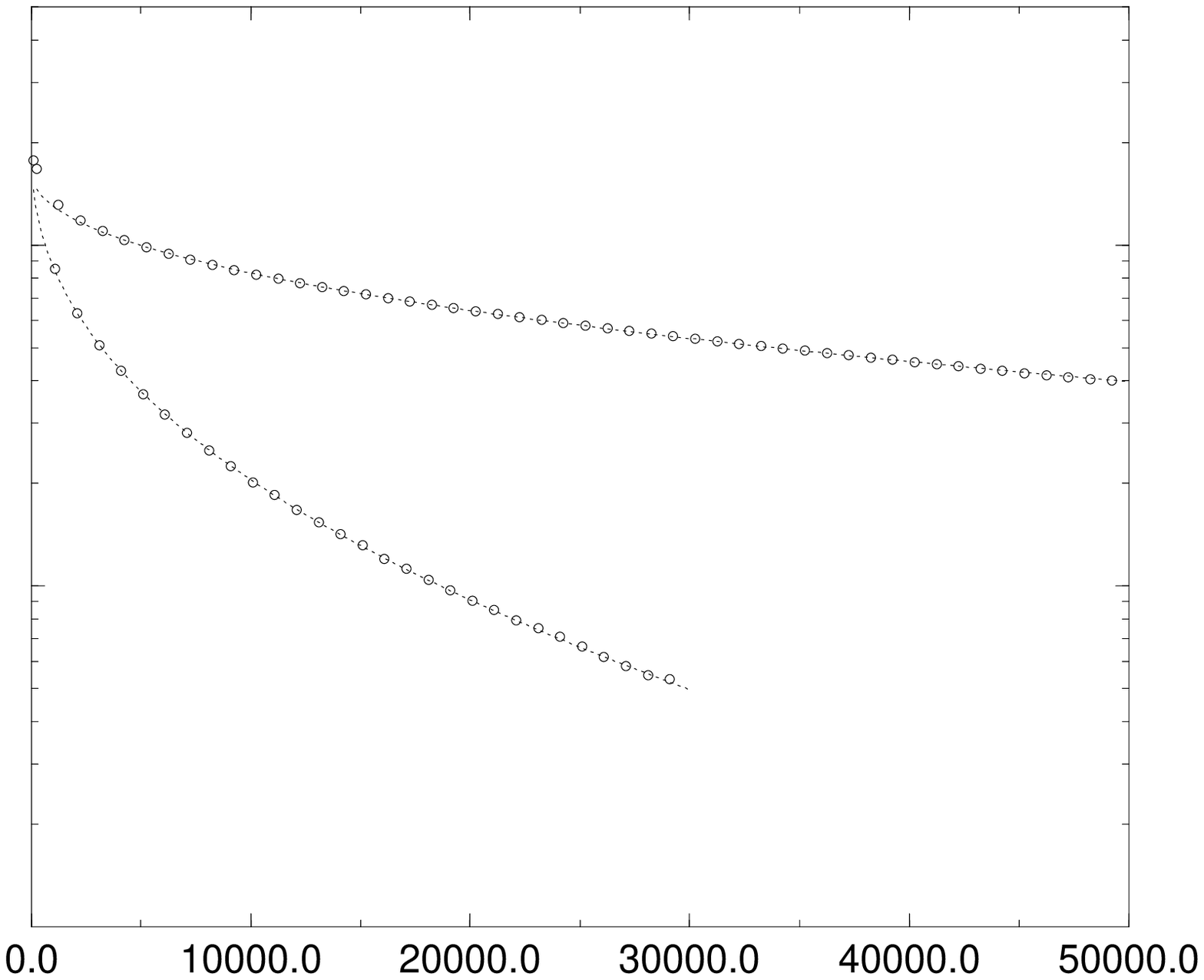}}}
\end{picture}
\caption{The auto-correlation 
in semi-log scale with $h=0.40$ (a) at the couplings
$J=0.10$, $0.20$, $0.30$ and $0.40$;
 (b) at the couplings
$J=0.50$, $0.60$ (from below). The dotted lines represent
 the stretched exponential form fitted to the curves.}
\label{f1}
\end{figure}

\begin{figure}[p]\centering
\epsfysize=12cm
\epsfclipoff
\fboxsep=0pt
\setlength{\unitlength}{1cm}
\begin{picture}(13.6,12)(0,0)
\put(0,-0.5){{\epsffile{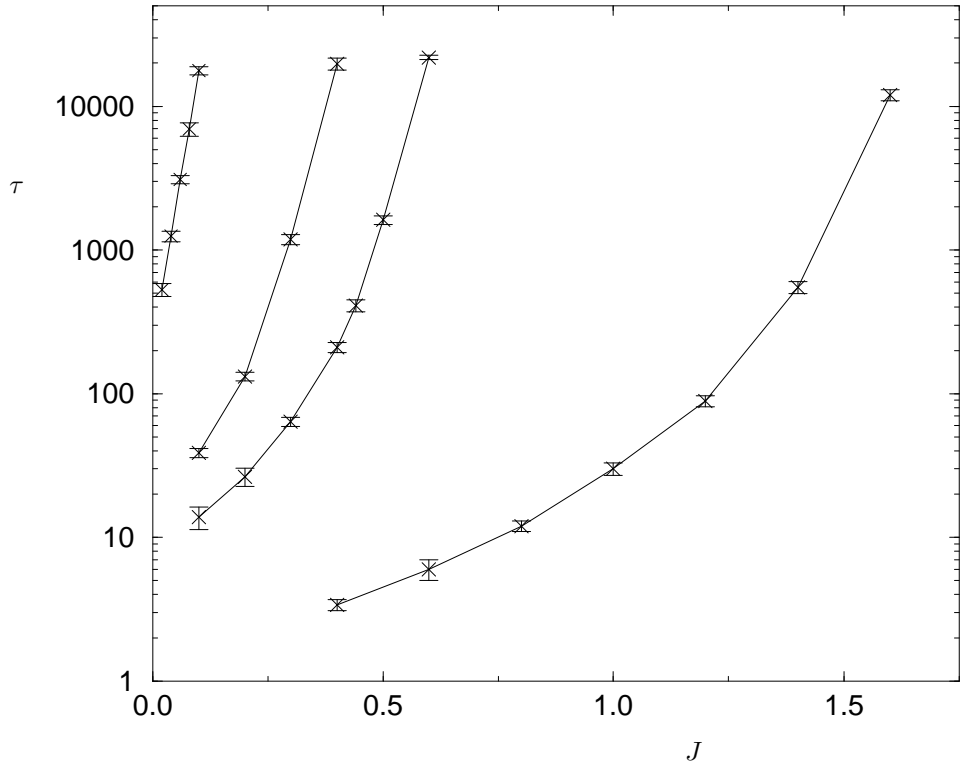}}}
\put(0.5,8.0){\makebox(0,0){\footnotesize $\tau$}}
\put(9.5,0.5){\makebox(0,0){\footnotesize $J$}}
\end{picture}
\caption{ The dependence of the correlation time $\tau$
on the coupling constant $J$. 
The corresponding fields are $h=0.80$, $0.50$, $0.40$ and $0.20$
respectively (from left). All curves show a non-Arrhenius behavior. 
}
\label{ftau}
\end{figure}

In the low temperature regime, the original Fredrickson Andersen model 
gives rise to a drastical enhancement of the relaxation time which is 
characterized inevitably with glassy materials. As demonstrated in 
\cite{schutri} there is no indication for a real glass transition or a critical 
temperature as predicted by mode--coupling theory \cite{Goetze1,Goetze2}. 
For large time $t$, empirical approaches  
suggest a stretched exponential decay of 
the auto-correlation function 
\begin{equation}
 A(t) \sim  exp[-(t/\tau)^\gamma].  
\label{esho40}
\end{equation}
where the exponent $\gamma$ is presumable not an universal exponent,
i.e. weakly depending on $h$.  
As a function of the $h$ (inverse temperature),
the relaxation time $\tau$ increases faster than according to an 
exponential law $\ln \tau \sim c+h$ manifested as a non-Arrhenius behavior
but there is no singularity $\tau \rightarrow \infty$ at finite temperatures
as suggested by the Williams--Landel--Ferry relation \cite{WLF}. 
The exponent $\gamma$ offers a weak dependence
on $h$ which is also confirmed by our simulations presented below. 
For the generalized SFM$\left[ 2,2\right] $ 
we are interested in the role of the extra coupling $J$. 
Physically, 
a positive exchange coupling $J > 0$ and $h > 0$ intend to support 
the creation of solid--solid pairs which are partially frozen in, e.g. 
such a coupling tends to enhance the relaxation time.\\  
We observe that for fixed $h$, even if it is small, the auto-correlation 
decays in a stretched exponential form for large time $t$
and more interestingly, the relaxation time $\tau$
increases also rapidly following a non-Arrhenius law
when the coupling $J$ increases.
The exponent $\gamma$ exhibits also a weak dependence
on the coupling $J$.
In Fig. \ref{f1}, the auto-correlation for
$h=0.40$ and for different values of the coupling $J$
is displayed in semi-log scale with lines of circles.
Obviously the decay is not an exponential one.
The dotted lines are the stretched exponential fit
to the curves. We see clearly the fit is rather good. 
The resulting relaxation time $\tau$ and the exponent
$\gamma$ for different $h$ and $J$
are listed in Table \ref {t1}.
For comparison, results for the original Fredrickson Andersen model
($J=0$) are also included.
In Fig. \ref {ftau}, we have plotted the correlation time 
$\tau$ as a function of the coupling $J$ for different
values of $h$ in semi-log scale.
Clearly it is a non-Arrhenius behavior.

\begin{figure}[h]
\epsfysize=6.5cm
\epsfclipoff
\fboxsep=0pt
\setlength{\unitlength}{0.6cm}
\begin{picture}(9,9)(0,0)
\put(-1.,-0.5){{\epsffile{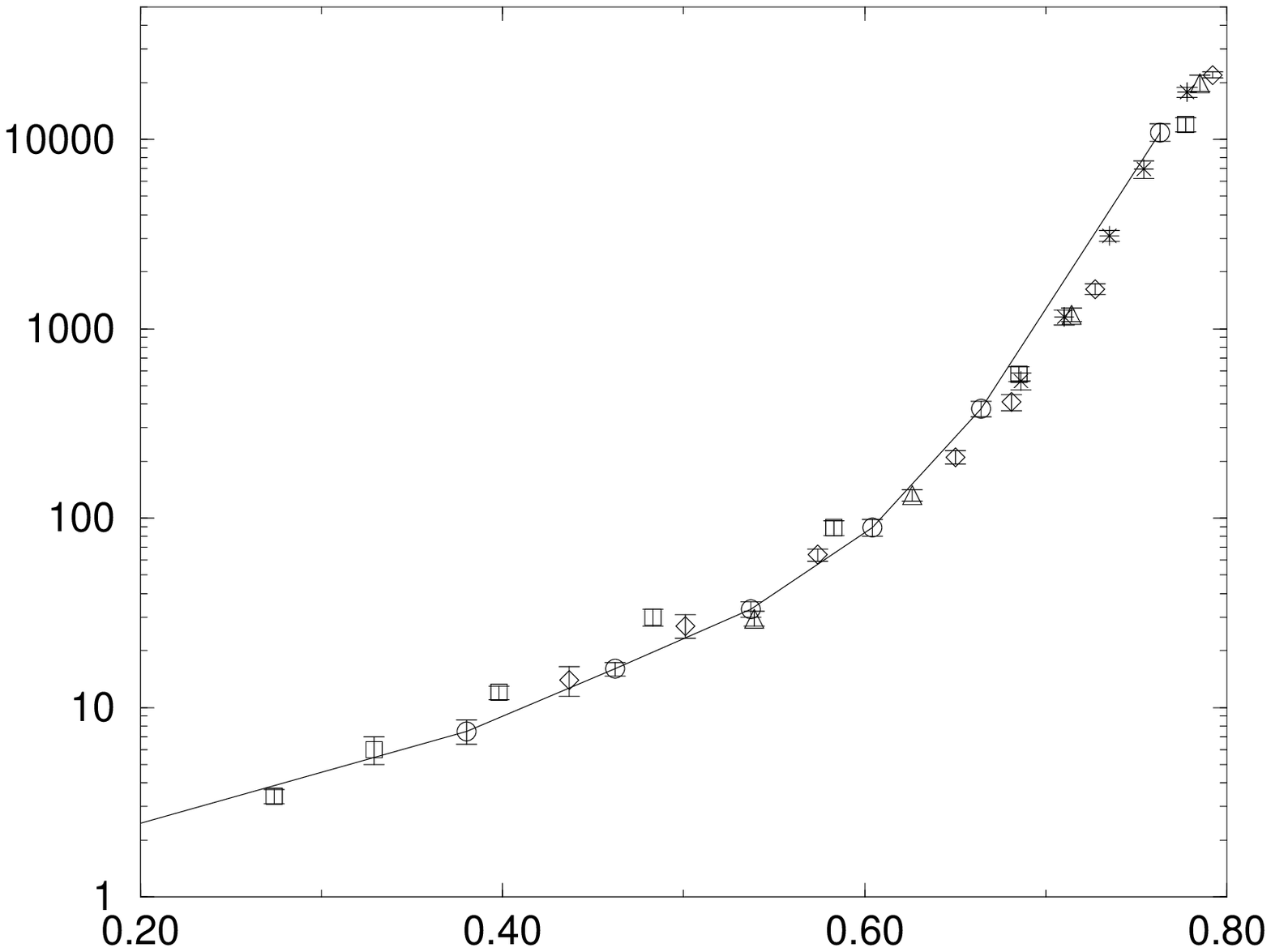}}}
\put(0.0,7.5){\makebox(0,0){\footnotesize $\tau$}}
\put(9.5,0.5){\makebox(0,0){\footnotesize $M$}}
\epsfysize=6.5cm
\put(12.,-0.5){{\epsffile{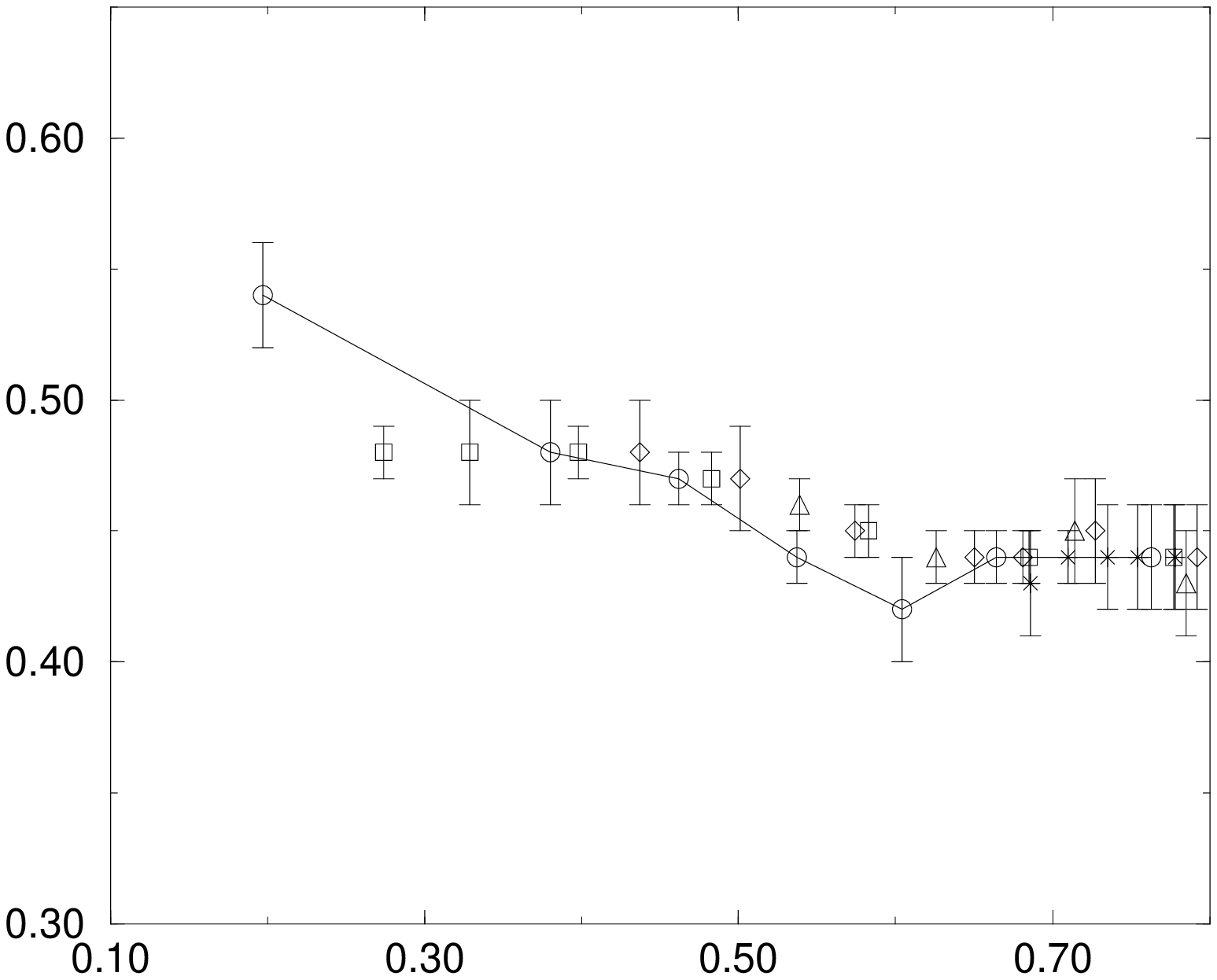}}}
\put(13.0,7.5){\makebox(0,0){\footnotesize $\gamma$}}
\end{picture}
\caption{The collapse of (a) correlation times $\tau$
in semi-log scale and (b) the exponent $\gamma$.
Circles with a solid line, squares, diamonds, triangles and stars
correspond to $J=0.00$, $h=0.20$, $h=0.40$ , $h=0.50$ and $h=0.80$
respectively. Data are taken from the first five blocks
of Table \protect\ref {t1}.}
\label{f2}
\end{figure}

Our model has two parameters $h$ and $J$.
For large $h$ and/or $J$, a strong freezing process manifested in a 
strong slowing down of the relaxation time is observed and
the auto-correlation shows similar dependence on both
$h$ and $J$, respectively. Alternatively to the autocorrelation function 
let us consider the magnetization $M(h,J)$. This quantity is an essential 
one also within our re-interpretation of the kinetic Ising model as an 
appropriate 
candidate to describe glasses. The magnetization corresponds to the 
density of the immobile particles.
In Fig. \ref {f2}, the dependence of the relaxation time $\tau$
and the exponent $\gamma$ on $M(h,J)$ are depicted for different
couplings $J$ and different fields $h$. A nice collapse of the data
is observed. A non-zero coupling $J$ practically induces
a short range spatial correlation. 
However, our results show that such a short spatial correlation length
does not change dramatically the properties of the glass system.

\begin{figure}[h]
\epsfysize=6.5cm
\epsfclipoff
\fboxsep=0pt
\setlength{\unitlength}{0.6cm}
\begin{picture}(9,9)(0,0)
\put(-1.,-0.5){{\epsffile{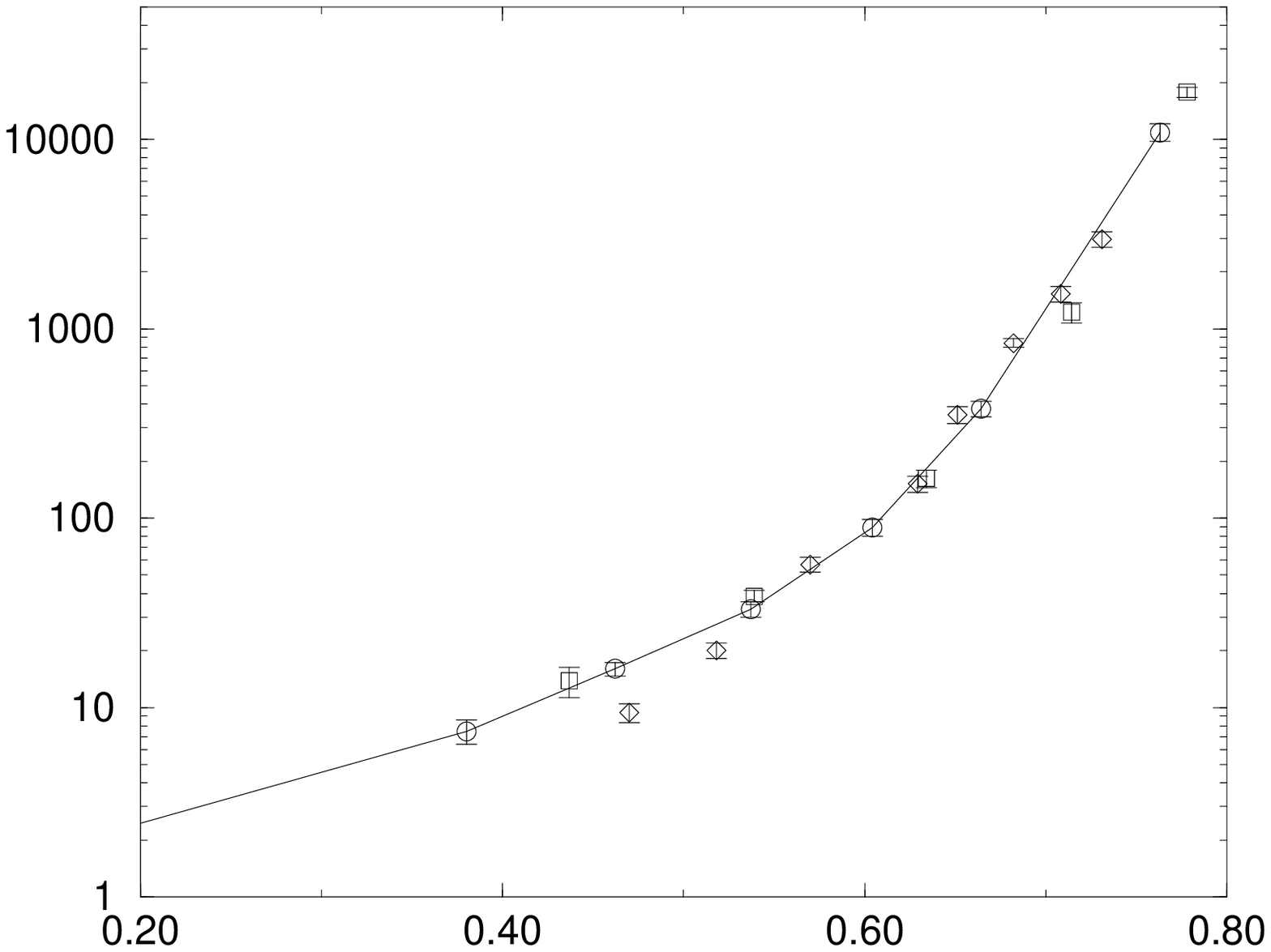}}}
\put(0.0,7.5){\makebox(0,0){\footnotesize $\tau$}}
\put(9.5,0.5){\makebox(0,0){\footnotesize $M$}}
\epsfysize=6.5cm
\put(12.,-0.5){{\epsffile{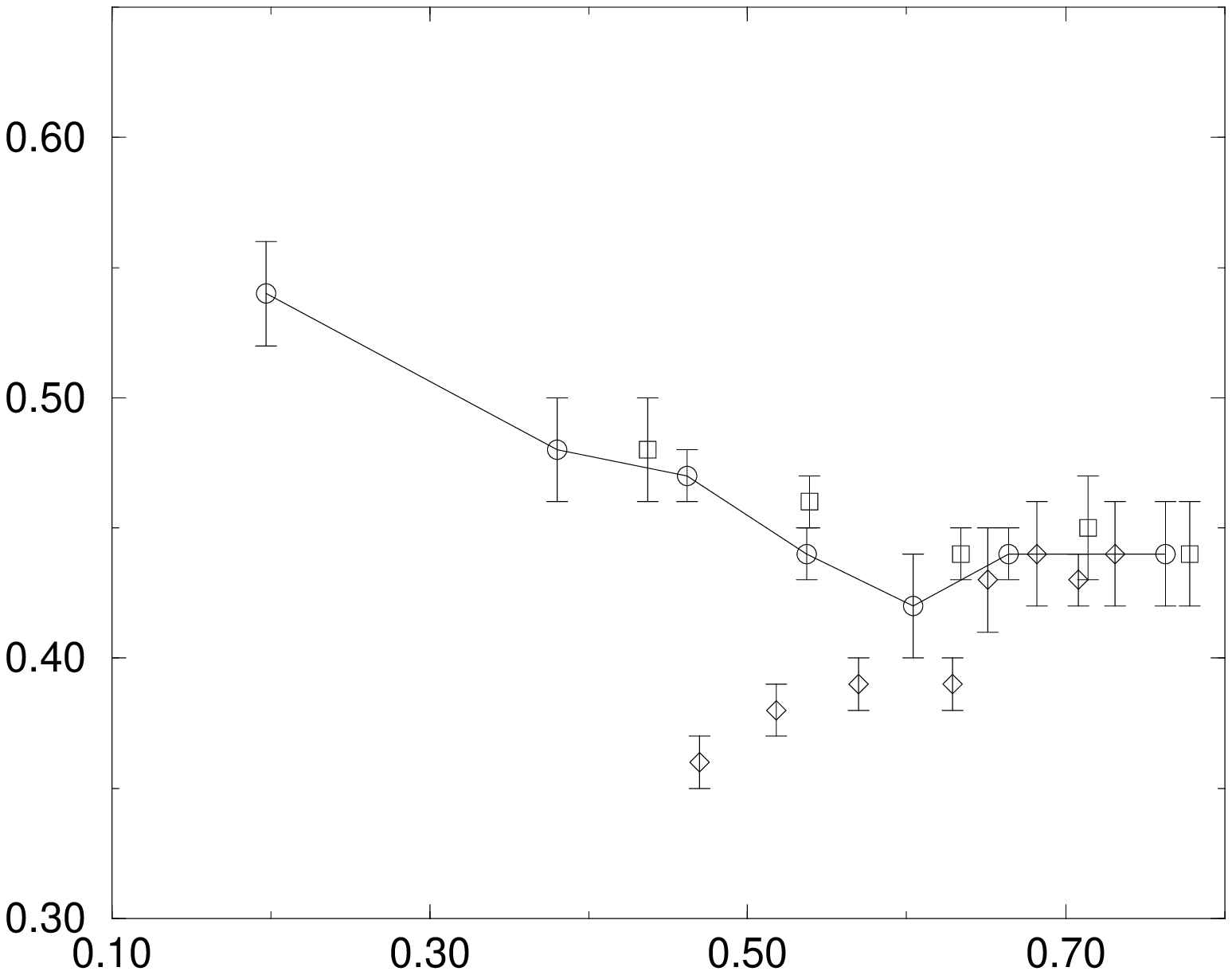}}}
\put(13.0,7.5){\makebox(0,0){\footnotesize $\gamma$}}
\end{picture}
\caption{The collapse of (a) correlation times $\tau$
in semi-log scale and (b) the exponent $\gamma$.
Circles with a solid line, squares and diamonds
correspond to $J=0.00$, $J=0.10$ and $h=1.00$ with negative $J$
respectively. Data are taken from the last two blocks
of Table \protect\ref {t1}.}
\label{f3}
\end{figure}

The interaction can be also anti-ferromagnetic, 
i.e. with a negative coupling constant $J$.
In this case, the relaxation time $\tau$ decreases
when the magnitude of $J$ becomes larger.
This can be seen in the last block of Table \ref {t1}.
However, we again find a nice collapse of th data
with negative $J$ and positive $J$ when the magnetization
$M$ is chosen to be the scaling variable.
This is shown in Fig.\ref {f3} (a).
The situation is slightly different for the exponent 
$\gamma$. In Fig. \ref {f3} (b), 
for small $|J|$, it joins to the data points
with positive $J$. But for bigger $|J|$, i.e. small
$M$ and with short
relaxation times $\tau$, the exponent $\gamma$ decreases
rather than increases as in the case of positive $J$.
This phenomenon is also understandable. A static 
antiferromagnetic coupling favors the coexistence of both 
liquid and solid like regions in the neighborhood.
But the limit of $J \rightarrow -\infty$ for fixed $h$
does not exactly correspond to a high temperature state.

\section{conclusions}

We obtain as a main result that the dynamics of the 
generalized SFM$\left[ 2,2\right]$ is controlled only by the 
magnetization $M(h,J)$, i.e.
the density of up (or down) spins. 
When we plot the correlation time $\tau$ as a function
of $M(h,J)$, all data points for different
values of $h$ and $J$ collapse to a single curve.
The collapse of the data points for the exponent
$\gamma$ is also observed except for the case with a strong
antiferromagnetic coupling.
These results show that the stretched exponential 
behavior is rather universal.

Our simulations have not yet covered the regime near the
critical point of the standard Ising model ($h \rightarrow 0$
but $hJ$ remains finite at its critical value).
In this critical regime, both the glass transition and
the second order phase transition take place.
To study the mixed critical behavior of two phase transitions
is an interesting extension of the present work.

Finally, it should be remarked that the empirical stretching exponent
$\gamma$ is rather close to 0.5. This result is in agreement with
recent analytical calculations \cite{schutri}.

\begin{table}[h]\centering
\begin{tabular}{ccc|ccc}   
 \hline
$(h,J)$    & $\tau$        & $\gamma$  & $(h,J)$     & $\tau$        & $\gamma$ 
\\
(0.2,0.0)   &  2.4(4)      &  0.54(2)  & (0.2,0.4)   & 3.4(3)       & 0.48(1)\\
(0.4,0.0)   &  7.5(11)     &  0.48(2)  & (0.2,0.6)   & 6.(1)        & 0.48(2)\\
(0.5,0.0)   &  16.(1)      &  0.47(1)  & (0.2,0.8)   & 12.(1)       & 0.48(1)\\
(0.6,0.0)   &  33.(3)      &  0.44(1)  & (0.2,1.0)   & 30.(3)       & 0.47(1) \\
(0.7,0.0)   &  89.(9)      &  0.42(2)  & (0.2,1.2)   & 89.(8)       & 0.45(1)\\
(0.8,0.0)   & 378.(36)     &  0.44(1)  & (0.2,1.4)   & 550.(52)     & 0.44(1)\\
(1.0,0.0)   & 10900.(1145) &  0.44(2)  & (0.2,1.6)   & 12000.(1040) & 0.44(2)\\
\hline
\hline
 $(h,J)$    & $\tau$        & $\gamma$  & $(h,J)$     & $\tau$        & 
$\gamma$\\
(0.4,0.1)   &  14.(3)      &  0.48(2)  & (0.5,0.1)   & 39.(3)       & 0.46(1)\\
(0.4,0.2)   &  26.(4)      &  0.47(2)  & (0.5,0.2)   & 132.(9)      & 0.44(2)\\
(0.4,0.3)   &  64.(5)      &  0.45(1)  & (0.5,0.3)   & 1185.(98)    & 0.45(2)\\
(0.4,0.4)   & 210.(17)     &  0.44(1)  & (0.5,0.4)   & 19800.(1940) & 0.43(2) \\
(0.4,0.5)   &1620.(110)    &  0.45(2)  &    &       & \\
(0.4,0.6)   &21900.(800)   &  0.44(2)  &    &     & \\
\hline
\hline
 $(h,J)$    & $\tau$        & $\gamma$  &    &        & \\
(0.8,0.02)   &  530.(54)      &  0.43(2)  &    &        & \\
(0.8,0.04)   &  1250.(105)    &  0.44(1)  &    &     & \\
(0.8,0.06)   &  3090.(201)    &  0.44(2)  &    &     & \\
(0.8,0.08)   & 6950.(747)     &  0.44(2)  &    &  &  \\
(0.8,0.10)   &17750.(1140)    &  0.44(2)  &    &       & \\
\hline
\hline
 $(h,J)$    & $\tau$        & $\gamma$  & $(h,J)$      &  $\tau$      & 
$\gamma$\\
(0.4,0.1)   &  14.(3)      &  0.48(2)  & (1.0,$-$0.02)  &  2963.(274) & 0.44(2) 
\\
(0.5,0.1)   &  39.(3)      &  0.46(1)  & (1.0,$-$0.06)  &  840.(45)   & 
0.44(2)\\
(0.6,0.1)   &  162.(17)    &  0.44(1)  & (1.0,$-$0.10)  &   152(15)   & 
0.39(1)\\
(0.7,0.1)   & 1220.(150)   &  0.45(2)  & (1.0,$-$0.14)  &  57.(5)     & 
0.39(1)\\
(0.8,0.1)   &17750.(1140)  &  0.44(2)  & (1.0,$-$0.18)  &   20(2)     & 
0.38(1)\\
            &              &           & (1.0,$-$0.22)  &   9.(1)     & 
0.36(1)\\
\hline
\end{tabular}
\caption{The correlation time $\tau$ and the exponent $\gamma$
measured for different different couplings $J$ and $h$.
}
\label{t1}
\end{table}

\newpage

\newpage

\newpage

\end{document}